\documentclass[a4paper,fleqn,authoryear]{cas-sc}

\usepackage[authoryear]{natbib}
\usepackage{graphicx}

\def\tsc#1{\csdef{#1}{\textsc{\lowercase{#1}}\xspace}}
\tsc{WGM}
\tsc{QE}

\begin{document}
\let\WriteBookmarks\relax
\def\floatpagepagefraction{1}
\def\textpagefraction{.001}

\shorttitle{Hoffman et al.}    

\shortauthors{}  

\title [mode = title]{Online Seakeeping Forecast Coupling System Identification With Nonlinear Pseudo Spectral Potential-Flow}  



\author[1]{Z.\ Hoffman}

\credit{}

\affiliation[1]{organization={RMIT University},
            city={Melbourne},
            postcode={3000}, 
            state={VIC},
            country={Australia}}

\author[1]{S.\ Vahaji} [orcid=0000-0001-6047-2163]
\cormark[1]


\ead{sara.vahaji2@rmit.edu.au}


\credit{}

\author[1]{A.\ Das}

\author[1]{M.\ Candon}

\author[2]{J.\ Nirman}

\author[1]{P.\ Marzocca}

\affiliation[2]{organization={Navantia Australia},
            addressline={}, 
            city={Melbourne},
            postcode={3000}, 
            state={VIC},
            country={Australia}}

\cortext[1]{Corresponding author}




\begin{abstract}
This paper presents a grey-box framework for a robust deterministic, real-time prediction of coupled surge--heave--pitch motion in waves. The approach combines a physics-based online system identification with a nonlinear pseudo-spectral potential flow (PSPF) solver. Rather than prescribing a complete hydrodynamic model from captive model tests or relying on a black-box surrogate, three coupled compact, dynamics equations are identified with constrained ridge-regression of a short time-history of free-running body motion data in waves. The PSPF solver is then used to provide the future incident wave at the body's location. The PSPF and dynamic system models are coupled for continuous real-time future motion prediction in a receding horizon architecture, while the model coefficients are re-identified online to enable robust prediction to changing wave-conditions. This novel framework is demonstrated based on CFD data that has been validated for a freely floating two-dimensional rectangular box in waves. The results demonstrate a feasible framework for accurate motion forecasting up to around eight wavelengths into the future utilising data that can be gathered onboard a ship in real operations. Furthermore, it is shown that the framework is robust to changing sea-states and agnostic to underlying model geometry.
\end{abstract}


\begin{highlights}
\item Linear and Nonlinear system identification for 3DOF seakeeping equations from 10 wave periods of free running motion data
\item PSPF wave and identified body dynamic models are coupled for continuous online future motion prediction
\item Online identification of coefficients enables model robustness to changing environment
\item Future motion predictions shown to be faster then real time
\end{highlights}



\maketitle

\section{Introduction}
Accurate, real-time deterministic predictions of ship sea-keeping is important for ship design, control, real-time decision support, and increasingly digital-twinning applications. In particular \cite{Hasan2024} emphasise the importance of interpretable, data-driven model discovery to support autonomous-ship digital twins. 

In creating such a model we might look first towards first principle 'white-box' models such as a transient naiver stokes solver (CFD) or potential flow solver (PF). Whilst CFD can resolve strongly non-linear wave--body interaction, viscous effects, and large motions, it's computational cost remains prohibitive for repeated, long-duration, or real-time simulations. Classical sea-keeping tools based on strip theory or linear potential flow are much more efficient, but their accuracy deteriorates when the assumptions of linearity, small amplitude waves and motions, and lack of viscous effects are violated. For example, \cite{Ozbulut2022} showed that quadratic damping and other nonlinear viscous effects materially influence the hydrodynamic coefficients of a 2-D body in roll. \cite{Zhang2025} showed that second-order wave loads are important for manoeuvring prediction in waves, while \cite{Sulisz1993} demonstrated that second-order diffraction loads on a semi-submerged rectangular box are significant and even exceed the corresponding first-order loads.

A common path to interpretable surrogate models is to assume a parametric structure and identify coefficients from virtual or experimental captive model tests. Such models for sea-keeping include the Cummins equation \cite{Newman1977} \cite{Davidson2015} \cite{Lu2022}, and the non-linear Fossen marine craft model \cite{Fossen2002}. 
\cite{Alexanderson2022} lists techniques taken by researchers and engineers for recovering the coefficients of a parametric model including; Kalman Filter (KF), constrained Least Squares (LS), and Support Vector Regression (SVR). The work of \cite{Alexanderson2022} also emphasises the importance model-selection, data-quality, multi-colinearity, and measurement noise affecting identified coefficients.  

Recently, sparse-regression has been used to expand system identification beyond fixed model structures. \cite{Shi2020} proposed a data-driven parametric identification framework for estimating coefficients that relate the state space to the excitation force on a wave energy converter using sparse regression, with gaussian process (GP) for uncertainty estimates. \cite{Liu2024} introduced a hydrodynamic-dictionary-based identification framework (HLAR) that builds candidate terms from classical maneuvering models and uses sequential thresholded least squares with Bayesian optimisation to tune sparsity hyperparameters, demonstrating accurate, low-complexity models across ship and Autonomous Unmanned Vessel (AUV) datasets. For long-horizon maneuvering prediction, \cite{Tian2026} further highlight how over-parameterised mechanistic identification can diverge, and propose a progressive optimised identification framework to improve stability and long-term predictive capability for data-limited scenarios.

In creating a parametric model of a floating-body wave-energy converter \cite{Davidson2015} utilised the Cummins framework and showed that radiation dynamics can be represented in compact state-space form, with linear hydrodynamic parameters identified from numerical wave tank data. Using this framework, they were able to produce linear models that are more representative across operating amplitudes than classical small-amplitude potential-flow coefficients alone. 

Nonlinear phase-resolved wave models, coupled with linear response models, have also been previously investigated for deterministic short-term ship motion forecasting. In particular, PSPF formulations are attractive because they provide a white-box methodology that can propagate highly nonlinear waves measured from ship X-band radar measurements \cite{Blondel-Couprie2012}, and when coupled with a linear model as in \cite{Fucile2016} or with a Cummins style model as in \cite{Wang2025} they have shown to provide accurate medium term motion forecast.

Increasingly common are nonparametric approaches, often black-box models with little prior physical assumption and constraints. \cite{Ouyang2023} proposed a local Gaussian Process Regression (GPR) framework that reduces the cost of classical GPR while maintaining satisfactory accuracy. \cite{DAgostino2022} compared different reccurent type neural networks including Long Short Term Memory (LSTM) trained on CFD data with far field wave probes as input to predict vessel motions in extreme sea states, while \cite{Silva2022} similarly developed an LSTM-based surrogate for 6-DoF ship motion in waves and showed that such models can represent free-running ship responses accurately for specific operating conditions. While these methods demonstrate strong capabilities for very short-term motion forecasting, they generally do not recover an explicit identified equation structure or directly identify interpretable hydrodynamic coefficients. Consequently, the capacity of these black-box approaches for robust generalisation and long-term forecasting accuracy remains questionable.

These developments collectively indicate an ongoing drive to develop more robust and accurate models that (i) enable real-time prediction for online use, (ii) are robust through preserving physical interpretability, and (iii) are valid to capture real-world non-linearities.

\subsection{Novelty and Contribution}
The principal contribution of this work is a grey-box framework for, real-time hydrodynamic prediction that couples an interpretable, physics based, dynamic models with a nonlinear incident-wave forecast in an online receding-horizon architecture. The potential-flow solver provides the predicted incident free-surface elevations at multiple points along the body's location, which are then used as forcing inputs for both system identification and numerical integration of the identified dynamic equations.

Rather than the more costly and less robust approach of predetermined linear transfer functions from captive experimental model tests or a purely black-box surrogate, constrained ridge regression is used to identify surge, heave and pitch coefficients from a short period of free running motion and wave data. A novel online re-identification procedure updates the model coefficients as new data comes in, allowing the identified equation to remain accurate to changing sea-conditions and even model geometries.  

A comparison was made between the identification of linear only coefficient terms and a nonlinear model where the basis is automatically determined though  sparse identification of nonlinear dynamics (SINDy). It was found that a model with linear only coefficients achieved similar test performance to a model with nonlinear coefficent terms, but the linear model remains preferable due to the stability of the identification and numerical integration. 

\section{Canonical Problem}
The canonical test case considered in this study is a floating rectangular box undergoing coupled surge, heave, and pitch motion in irregular waves. 

\subsection{Numerical Verification}

The numerical simulations were performed using the commercial URANS solver Ansys Fluent 2025R2. The CFD setup employed explicit volume-of-fluid free-surface capturing, dynamic overset meshing, and the \(k\text{--}\omega\) SST turbulence model with a maximum wall y+ of 23. Spatial discretisation was based on a structured Cartesian background mesh. The computational domain had a total length of \(96.7 \,\mathrm{m}\), including a numerical beach of length \(30.8\,\mathrm{m}\) to reduce wave reflections at the computational domain outlet boundary. The water depth was set to \(1.50 \,\mathrm{m}\), and the water density was taken as \(\rho_{water}=998.2\,\mathrm{kg/m^3}\).

\begin{figure}
    \centering
        \includegraphics[width=0.5\linewidth]{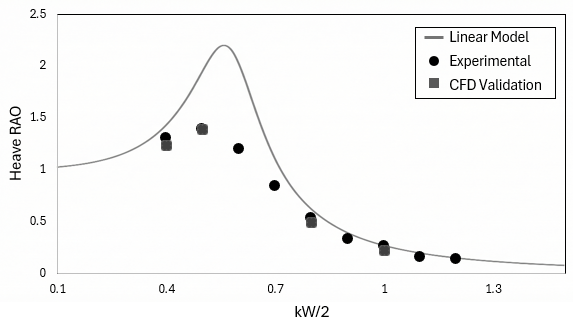}
    \caption{Response amplitude operator for a captive heave model of a box in regular waves validation against Rodrigues et al \cite{Rodriguez2016}. $\frac{kW}{2}$ is the wave number $k$ multiplied by half of the box width $W$.}
    \label{fig:RAO}
\end{figure}

\begin{figure}
    \centering
    \begin{minipage}{0.45\textwidth}
        \centering
        \includegraphics[width=\linewidth]{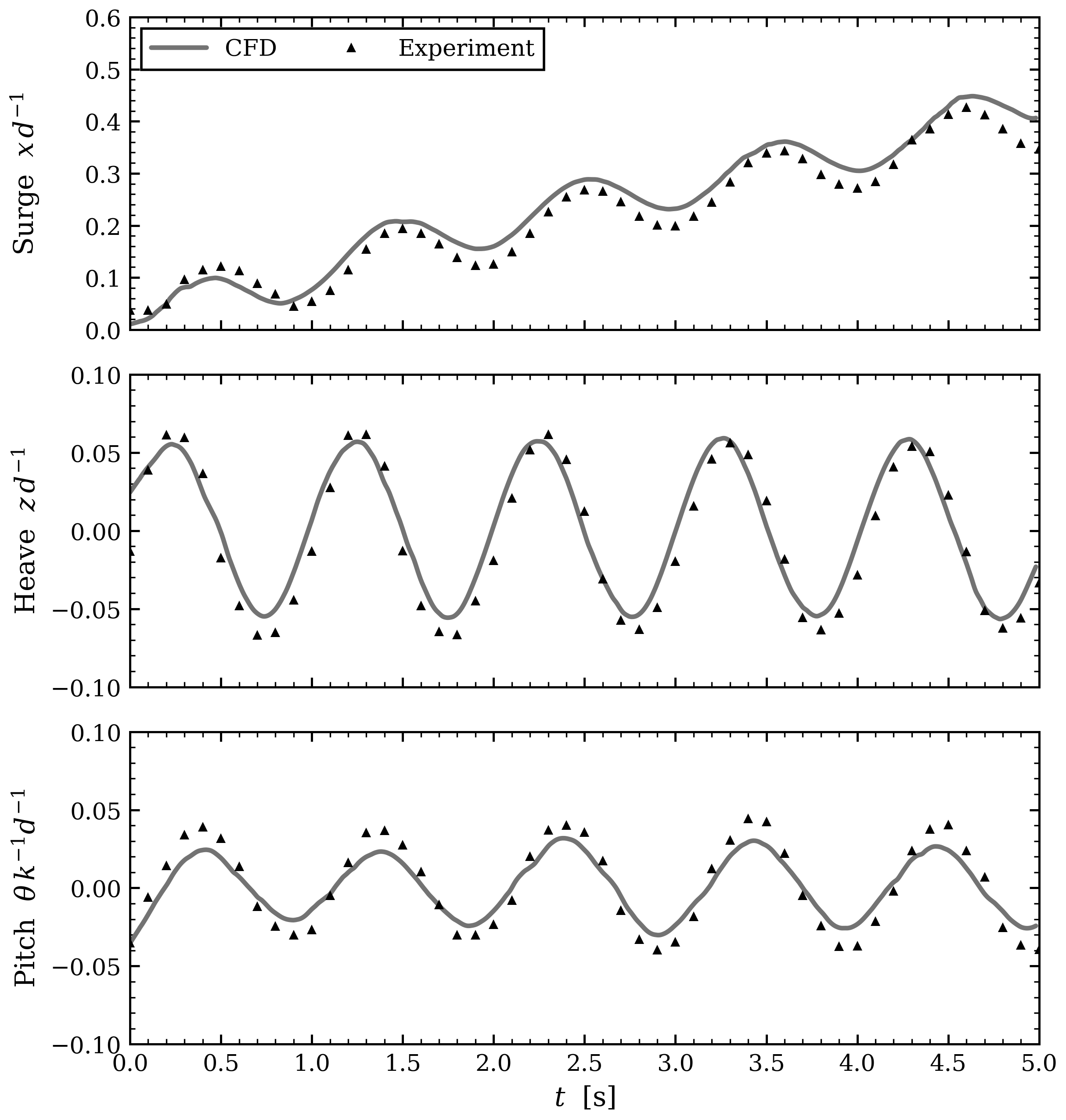}
        \label{fig:Coordinate-sys}
    \end{minipage}
    \hfill
    \begin{minipage}{0.45\textwidth}
        \centering
        \includegraphics[width=\linewidth]{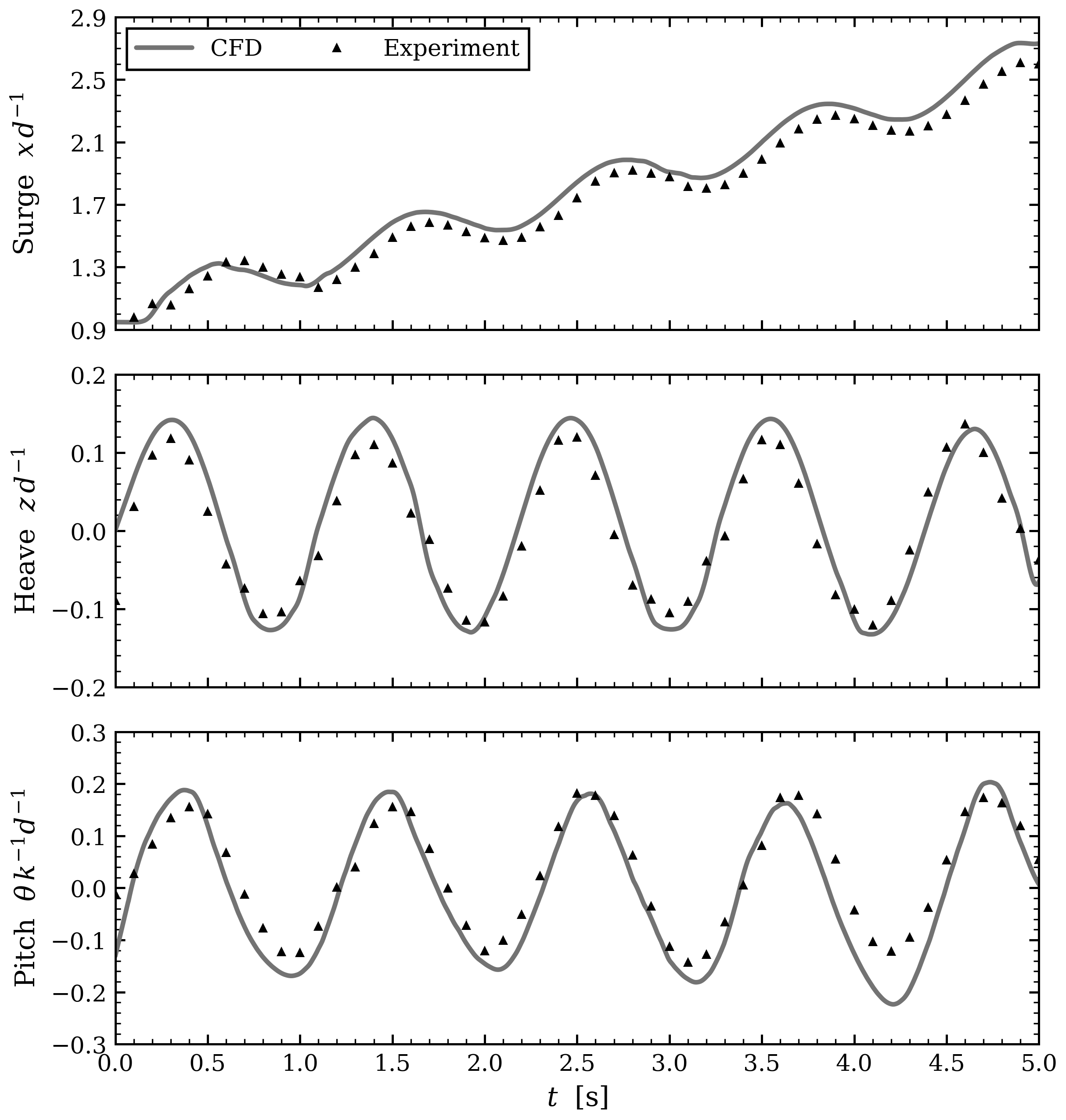}
    \end{minipage}
    \caption{Validation of two-dimensional box three-degree of freedom response to regular wave against \cite{Ren2015} for (a) $T=1.2\text{s}$, $\eta_i=0.05\text{m}$ and (b) $T=1.2\text{s}$, $\eta_i=0.02\text{m}.$}
    \label{fig:3dof-valid}
\end{figure}

The CFD numerical settings were validated against available experimental data from \cite{Rodriguez2016} and \cite{Ren2015}. The predicted response-amplitude operator (RAO) for a two-dimensional box captive heave responce to waves shows good agreement across both non-linear and linear response regimes in Figure~\ref{fig:RAO}. The significant reduction in heave RAO near the box's natural frequency is primarily due to vortex shedding around the corners of the box's geometry. Finally, the predicted coupled 3-DoF box motions compare well with experimental results, as shown in Figure~\ref{fig:3dof-valid}. These results demonstrate that the numerical model is able to resolve the principal hydrodynamic features required for the present study. Taken together, these results provide confidence in the CFD model and establish a robust numerical foundation for the development of the surrogate model presented in this paper.

\begin{table}[0.45\textwidth]
    \centering
    \begin{tabular}{l c}
    \hline
    Parameter & Value \\
    \hline
    Box width, \(W_{box}\) & \(0.3\,\mathrm{m}\) \\
    Box height, \(h_{box}\) & \(0.2\,\mathrm{m}\) \\
    Box mass, \(M_{box}\) & \(29.9\,\mathrm{kg}\) \\
    Metacentric height, \(GM\) & \(0.0613\,\mathrm{m}\) \\
    \hline
    \end{tabular}
    \caption{Geometric and inertial properties of the floating box.}
    \label{tab:box-properties}
\end{table}

\section{Model Input}
\label{sec:model input}
A central modelling decision concerns the definition of the wave input supplied to the data-driven system identification framework outlined in Section \ref{sec:SINDy}.

\subsection{Wave Representation}
\begin{figure}
    \centering
    \includegraphics[width=0.9\linewidth]{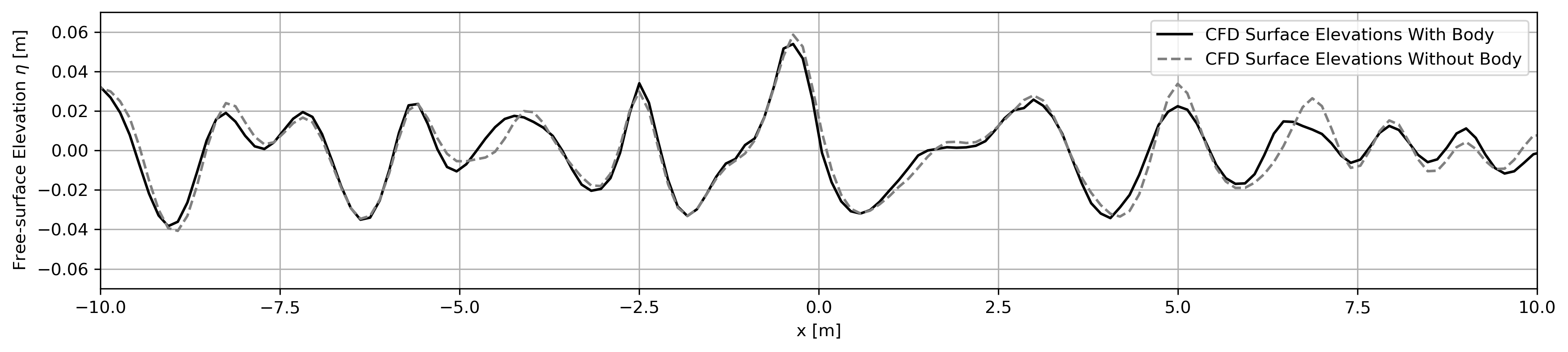}
    \caption{CFD surface elevation with vs without body present within the domain.}
    \label{fig:WH body vs no body}
\end{figure}

Under many simplified hydrodynamic problems including in linear potential-flow theory, linear sea keeping theory, and radiation-diffraction panel methods, the free-surface elevation in the presence of a floating body may be decomposed as
\begin{equation}
\eta_{\mathrm{tot}}
=
\eta_{\mathrm{I}}
+
\eta_{\mathrm{D}}
+
\eta_{\mathrm{R}},
\label{eq}
\end{equation}
where $\eta_{\mathrm{tot}}$ is the total surface elevation comprised of the undisturbed incident-wave elevation $\eta_{\mathrm{I}}$, the wave field diffracted by the body $\eta_{\mathrm{D}}$, and the wave field radiated by the body's motion $\eta_{\mathrm{R}}$. Despite it's validity not being exact for the nonlinear viscous CFD problem considered here, it was found to constitute a 'good' simplification justified by the overall effectiveness of the final model. Due, to the geometry of this specific 2D problem, diffraction effects are significant and can be seen in Figure \ref{fig:WH body vs no body} by comparing the equivalent surface elevations with no body present and the true surface elevations with the body there. 

\subsection{Consequence Of Incident Wave Input}
The identified model is sensitive to which of these wave-field representations is used as its input. A model trained using the body-modified surface elevation $\eta_{\mathrm{tot}}$ can reproduce the observed response accurately as the input already contains information generated by the body's presence and motion. However, it is computationally intractable to compute the fully nonlinear body-wave interaction in a non-empty domain in real time, especially under fluid-body motion coupling, viscous, and other non-linear fluid effects such as vortex shedding seen in Fig \ref{fig:3dof-valid}. But it is possible to, in real-time, propagate a fully non-linear incident wave-field upstream in an empty domain using a first principle solver described in Section \ref{sec:HOS-solver}.

For this reason, the surrogate model developed in the present study is driven exclusively by the equivalent undisturbed incident-wave field $\eta_{\mathrm{I}}$. The incident wave is propagated independently of the body's motion, while the effects of diffraction and radiation are represented implicitly in the identified motion equations explained in Section \ref{sec:SINDy}. The resulting model is therefore expected to remain most reliable within the range of wave conditions, body motions, and fluid-interaction regimes represented during identification. Extrapolation to substantially different wave steepnesses, frequencies, body geometries, or strongly nonlinear flow regimes requires additional training data which is considered in this code through continuously updating the model's coefficients online as explained in Section \ref{sec:online-training}.

\subsection{Input Vector}
To provide information on both the local wave elevation and its derivatives the incident field is predicted at the instantaneous fore and aft locations of the body. Multiple wave-height measurements was found to improve model robustness and accuracy by providing information related to the instantaneous wave slope and phase variation across the body, both of which are important for distinguishing the excitation of heave and pitch. For a box of width $b=0.30~\mathrm{m}$ and instantaneous horizontal position $x_{\mathrm{b}}(t)$, these inputs are defined as
\begin{align}
\eta_{1}(t)
&=
\eta_{\mathrm{I}}
\left(
x_{\mathrm{b}}(t)-\frac{B}{2},
t
\right),
&
\eta_{2}(t)
&=
\eta_{\mathrm{I}}
\left(
x_{\mathrm{b}}(t)+\frac{B}{2},
t
\right),
\label{eq:wave_input}
\end{align}

The derivatives, or the slope of the wave $\dot{\eta}_1$ and $\dot{\eta}_2$ are formulated with the central difference method described in Eq. \ref{eq:dx}, and their inclusion was found to add a significant amount of accuracy and robustness to the model.

The resulting wave-input for model training and deployment is then;
$$\eta_{1}(t) \qquad \dot{\eta}_{1}(t) \qquad \eta_{2}(t) \qquad \dot{\eta}_{2}(t) $$

\subsection{Pseudo-Spectral Potential Flow (PSPF) Governing Equations}
\label{sec:HOS-solver}
The work of~\cite{Klahn2021} presents a nonlinear pseudo-spectral potential-flow (PSPF) model capable of evolving the wave field. The numerical method is well suited to the present work as it provides a faster then real-time, highly efficient, white-box, nonlinear solver. The solver is re-implemented and modified for the application, with the numerical modifications detailed in Section~\ref{sec:solver-mods}.

Potential flow assumes an inviscid, incompressible and irrotational fluid, so that the velocity field can be written in terms of a scalar velocity potential $\Phi(x,z,t)$. Under these assumptions, the fluid motion is governed by the Laplace equation in the fluid domain,
\begin{equation}
    \frac{\partial^2 \Phi}{\partial x^2} +
    \frac{\partial^2 \Phi}{\partial z^2}
    = 0
    \qquad
    \text{for} \, -h<z<\eta(x,y,t)
    \label{eq:Laplace}
\end{equation}
with an impermeable flat seabed condition
\begin{equation}
    \frac{\partial \Phi}{\partial z}=
    0
    \qquad
    \text{for} \, z=-h.
    \label{eq:bottom-bc}
\end{equation}

The governing Zakharov formulation modifies the inviscid, irrotational and incompressible potential-flow equations by exploiting the fact that the system is uniquely determined by the free-surface elevation and the velocity potential evaluated at the free surface, $\Phi_s$, such that
$\Phi_s = \Phi|_{z=\eta}$,
\begin{equation}
    \Phi_s(x,t)=\Phi(x,\eta(x,t),t).
\end{equation}
The nonlinear free-surface boundary conditions form a pair of coupled nonlinear PDEs, Eq.~\eqref{eq: Kinematic BC} and
Eq.~\eqref{eq: dynamic BC}, that are advanced in time using an RK4 scheme.
\begin{equation}
    \frac{\partial \eta}{\partial t} =
    \left(1 +
    \left(\frac{\partial \eta}{\partial x}\right)^2\right)
    w_s -
    \frac{\partial \eta}{\partial x}
    \frac{\partial \Phi_s}{\partial x}
    \label{eq: Kinematic BC}
\end{equation}
\begin{equation}
    \frac{\partial \Phi_s}{\partial t} =
    -g \eta -
    \frac{1}{2}
    \left(\frac{\partial \Phi_s}{\partial x}\right)^2 +
    \frac{1}{2}
    \left(1 +
    \left(\frac{\partial \eta}{\partial x}\right)^2\right)
    w_s^2
    \label{eq: dynamic BC}
\end{equation}
where $w_s = \partial \Phi / \partial z|_{z=\eta}$ is obtained by solving the Laplace problem at each time step.

\subsubsection{Numerical Modifications}
\label{sec:solver-mods}

The reference implementation of ~\cite{Klahn2021} solves the fully three-dimensional problem in the horizontal $x$ and $y$ directions and vertical direction $s$. For the present 2D application, the governing equations were modified be removing the y-direction (Eq. \ref{eq: dynamic BC}, Eq. \ref{eq: Kinematic BC}, and Eq. \ref{eq:Laplace}). This also corresponds to the case of a unidirectional 3D long-crested wave field. Further modifications were made for mathematically equivalent reformulations of the original reference code designed to speed up the computation. For example, since $\eta$, $\Phi_s$ and $F$ are real, when a FFT or inverse-FFT is applied, only the non-negative real half of each spectrum is stored Eq. \ref{eq:k-modes}, halving the memory and arithmetic. 

For a $10.84$ s PF simulation relating to around 8 periods of wave data, for $2\times 248$ Fourier collocation points in the x-direction, and with a Dommermuth adjustment ramp of 5, and a vertical expansion order $N_s$ of $20$, the reference code is capable of computation in $10.108$ s. After the above mentioned modifications, the time was reduced to $0.8625$ s, leading to a speed up of $11.7$ times while the mean squared error between the reference and updated code is only $2\times 10^{-10}$. 

A new function was written to construct the initial Zakharov high-order spectral surface velocity potential based only on a free-surface elevation file. To reduce the required computation, the surface-elevation window read from file and the number of horizontal grid points $N_x$ solved for are scaled linearly with wavelength and prediction horizon, so that the solver need only include the upstream distance of waves capable of influencing the body within the prediction horizon.

The elevation file is read, centred, and mapped onto the solver grid using PCHIP interpolation. The file contains equally spaced surface elevation values spanning the domain, and the first value is appended at $x=L_x$ such that $\eta(L_x)=\eta(0)$, making the profile periodic. The Fourier coefficients of the surface elevation are then obtained from $\hat{\eta}=\frac{1}{2N_x}\mathrm{FFT}(\eta)$. Each Fourier coefficient $\hat{\eta}_m$ is associated with a Fourier mode $e^{ik_mx}$, where $m$ is the Fourier mode index and the corresponding wavenumber is
\begin{equation}
k_m=\frac{2\pi m}{L_x}
\qquad m=0,1,\ldots,N_x
\label{eq:k-modes}
\end{equation}
The surface elevation can therefore be represented as the superposition of positive real-valued Fourier modes obtained through the FFT, 
\begin{equation}
\eta(x)=\sum_{m=0}^{N_x}\hat{\eta}_m e^{ik_mx}
\end{equation}

The initial surface velocity potential $\Phi_s$ cannot be uniquely determined from a single wave-elevation snapshot alone. The initialisation assumes that every non-zero-mean component of the measured field consists of a linear wave travelling in the positive $x$-direction. Under the linear dispersion relation and deep water limit becomes,
\begin{equation}
    \omega^2_m = g |k_m| \mathrm{tanh}(|k_m| h)|
\end{equation}
Under the linear kinematic condition,
\begin{equation}
    \frac{\partial \hat{\eta}_m}{\partial t}
    =
    |k_m|\tanh(|k_m|h)\,\hat{\Phi}_{s,m}
\end{equation}
which, once rearranged, gives the relation used to extract $\Phi_s$ from the
known surface elevation $\eta$:
\begin{equation}
    \hat{\Phi}_{s,m} = -i \frac{\omega_m}{k_m \mathrm{tanh}(k_mh)} \hat{\eta}_m \qquad k_m>0
    \label{eq:phis-from-eta}
\end{equation}
Once $\hat{\Phi}_s$ is computed in the Fourier domain, it is converted back to physical space with the inverse Fourier transform. 

The method employed here will deviate from the true surface potentials in that it does not account for the physics of $-x$ wave components from body-diffraction or radiation. Direct information on the velocity potential can easily be extracted from CFD data, but, due to the constraints of the physical world, this cannot be done in practice aboard a vessel, where technologies such as wave radar enables the capability of only extracting the surface elevation field around a body.

\subsubsection{Potential Flow Code Verification and Benchmarking}
\label{sec:solver-verification}

\begin{figure}
\centering
    \includegraphics[width=0.9\linewidth]{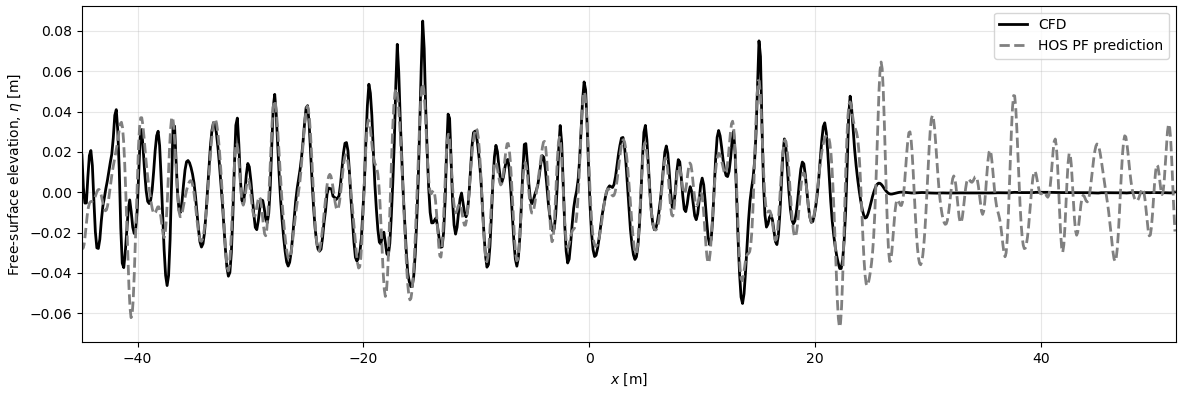}
    \caption{Comparison of the CFD and PF free-surface elevations approximately $8.3$ wave periods after initialisation across the full computational domain.}
    \label{fig:CFD-PF-comparison}

    \centering
    \includegraphics[width=0.9\linewidth]{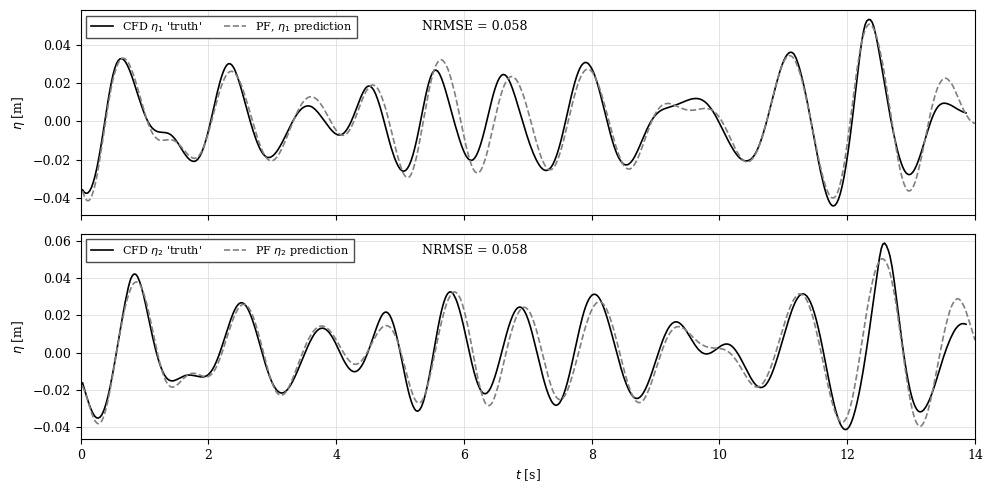}
    \caption{PF and CFD incident free-surface elevations at the fore and aft locations of the floating body.}
    \label{fig:WH-probes}
\end{figure}

The re-implementation of \cite{Klahn2021} PSPF reference code is verified to ensure that it maintains high solution accuracy when compared to a CFD 'truth' containing the equivalent wave field without a body present. Figure~\ref{fig:CFD-PF-comparison} compares the PF and CFD free-surface elevations across the full computational domain. The comparison is shown $10.0~\mathrm{s}$, after the PF solution was initialised from the CFD free-surface elevation. Near the upstream boundary, the solutions diverge because the PF domain is periodic, whereas the CFD domain uses an open-water prescribed wave inlet. Similarly, the downstream CFD domain contains a damping region that is not present in the PSPF. Agreement is therefore not expected across the complete domain. The strongest agreement occurs around the zero datum, corresponding to the region occupied by the floating body. 

The ability of the PSPF to propagate the incident wave field through the body region is quantified through comparing the predicted free surface elevations $\eta_{\mathrm{I},1}$ \& $\eta_{\mathrm{I},2}$ found with the empty CFD domain, shown in Figure~\ref{fig:WH-probes}. The NRMSE between the CFD and PSPF free-surface incident elevations prediction is $0.058$.

A refinement study was additionally carried out to determine the required horizontal resolution $N_x$, vertical expansion order $N_s$ and time-step $\Delta t$. Convergence was assessed from the predicted free-surface elevation and computational wall time, with the final discretisation selected to provide sufficient numerical convergence while retaining the computational speed required for online prediction.

\section{Surrogate Model Formulation}
\label{sec:SINDy}
This section presents the formulation of the proposed surrogate model for the coupled surge, heave, and pitch dynamics of the floating box. The particular combination of modelling assumptions, radiation-memory representation, and system identification strategy adopted here constitutes a novel contribution.

\subsection{Modelling Assumptions}
The surrogate model is formulated as a physics-based approximation of the Cummins equation for wave-body interaction. In compact form, the dynamics are written as
\begin{equation}
\left({M}+{A}_{\infty}\right)\ddot{{s}}
+{F}_{\mathrm{rad}}
+{F}_{\mathrm{hs}}
+{F}_{\mathrm{visc}}
=
{F}_{\mathrm{ext}}(\eta,\dot{\eta}),
\label{eq:rom_general}
\end{equation}
where \(\mathbf{s}=[x,z,\theta]^{\top}\) is the state vector of trajectories containing surge, heave, and pitch.

Here, \({M}\) denotes the rigid-body inertia matrix, \({A}_{\infty}\) is the infinite-frequency added-mass matrix, \({F}_{\mathrm{rad}}\) is the radiation force, \({F}_{\mathrm{hs}}\) represents hydrostatic restoring effects, \({F}_{\mathrm{visc}}\) denotes viscous damping, and \({F}_{\mathrm{ext}}\) is the wave-excitation force. 

The radiation contribution is written in Cummins form as
\begin{equation}
{F}_{\mathrm{rad}}
=
-\int_{0}^{t}{K}(t-\tau)\,\dot{\mathbf{s}}(\tau)\,\mathrm{d}\tau
\label{eq:radiation_force}
\end{equation}
where \({K}(t)\) is the retardation function. In the present work, this convolution is approximated using a state-space representation of the radiation-memory dynamics, such that the history-dependent radiation force is replaced by a set of extra ODE states that model a decaying sine wave. The implementation was inspired by the work of \cite{Davidson2015} and \cite{Lu2022}.

For a velocity \(\dot{\mathbf{s}}(t)\), the memory states satisfies the subsystem:  
\begin{equation}
\dot{y} = \dot{\mathbf{s}} - \sigma y - \omega w,
\qquad
\dot{w} = \omega y - \sigma w,
\label{eq:memory_generic_clean}
\end{equation}
where \(\sigma\) is a decay rate and \(\omega\) is an oscillation frequency. The magnitude of these hyperparameters are determined using the hyperparameter optimisation scheme further discussed in Section \ref{sec:Hyperamater-optimisation}.
\begin{align}
\dot{y}_x &= \dot{x} - \sigma_x y_x - \omega_x w_x,
&
\dot{w}_x &= \omega_x y_x - \sigma_x w_x,
\\
\dot{y}_{z z} &= \dot{z} - \sigma_{z z} y_{z z} - \omega_{z z} w_{z z},
&
\dot{w}_{z z} &= \omega_{z z} y_{z z} - \sigma_{z z} w_{z z},
\\
\dot{y}_{z \theta} &= \dot{\theta} - \sigma_{z \theta} y_{z \theta} - \omega_{z \theta} w_{z \theta},
&
\dot{w}_{z \theta} &= \omega_{z \theta} y_{z \theta} - \sigma_{z \theta} w_{z \theta},
\\
\dot{y}_{\theta z} &= \dot{z} - \sigma_{\theta z} y_{\theta z} - \omega_{\theta z} w_{\theta z},
&
\dot{w}_{\theta z} &= \omega_{\theta z} y_{\theta z} - \sigma_{\theta z} w_{\theta z},
\\
\dot{y}_{\theta \theta} &= \dot{\theta} - \sigma_{\theta \theta} y_{\theta \theta} - \omega_{\theta \theta} w_{\theta \theta},
&
\dot{w}_{\theta \theta} &= \omega_{\theta \theta} y_{\theta \theta} - \sigma_{\theta \theta} w_{\theta \theta}.
\end{align}

The hydrostatic restoring force is represented using,
\begin{equation}
{F}_{\mathrm{hs}}
=
{K}_{1}{s}
+{F}_{\mathrm{nl}}({s})
\label{eq:hydrostatic_force}
\end{equation}
where the linear terms represent the dominant restoring behaviour, and \({F}_{\mathrm{nl}}({X})\) denotes additional nonlinear restoring contributions. Hydrostatic terms relating to surge were set to zero.

The viscous damping contribution is written as
\begin{equation}
{F}_{\mathrm{visc}}
=
{C}_{1}\dot{{s}}
+\mathbf{C}_{2}\lvert \dot{{s}} \rvert \dot{{s}}
\label{eq:viscous_force}
\end{equation}
where the first term represents linear damping and the second term represents quadratic drag as per the Morrison equation.

The external wave forcing is assumed to depend on the local wave elevation and its time derivative, and is represented in the general form;
\begin{equation}
{F}_{\mathrm{ext}}
=
{B}_{1}\eta
+{B}_{2}\dot{\eta}
+{B}_{3}\eta {s}
+{B}_{4}\eta\dot{{s}}
\label{eq:excitation_force}
\end{equation}
allowing direct wave excitation and state-dependent wave-force coupling to be represented within the candidate library.

The measured time histories used for system identification are
$$
s(t) = x(t),\ \dot{x}(t),\ 
z(t),\ \dot{z}(t),\
\theta(t),\ \dot{\theta}(t) $$
$$
w(t) = \eta_{\mathrm{I},1}(t),\ \dot{\eta}_{\mathrm{I},1}(t), \, \eta_{\mathrm{I},2}(t),\, \dot{\eta}_{\mathrm{I},2}(t) $$
$$
m(t) = y_x(t),\, y_{zz}(t),\, y_{z\theta}(t),\, y_{\theta\theta}(t),\, y_{\theta z}(t),\ w_x(t),\, w_{zz}(t),\, w_{z\theta}(t),\, w_{\theta\theta}(t),\, w_{\theta z}(t)
$$

Using these signals, separate sparse models are identified for surge, heave, and pitch. The identified dynamics are written as
\begin{equation}
    \ddot{x} = \Theta_x {\Xi}_x \qquad
    \ddot{z} = \Theta_z {\Xi}_z \qquad
    \ddot{\theta} = \Theta_\theta {\Xi}_\theta 
\end{equation}
where \(\Theta_x\), \(\Theta_z\), and \(\Theta_\theta\) are the candidate libraries for surge, heave, and pitch, respectively, and \({\Xi}_x\), \({\Xi}_z\), and \({\Xi}_\theta\) are the corresponding coefficient vectors that describe the dynamic model ODEs. 

\subsection{Motion Derivatives}

\begin{figure}
\centering
\includegraphics[width=0.8\linewidth]{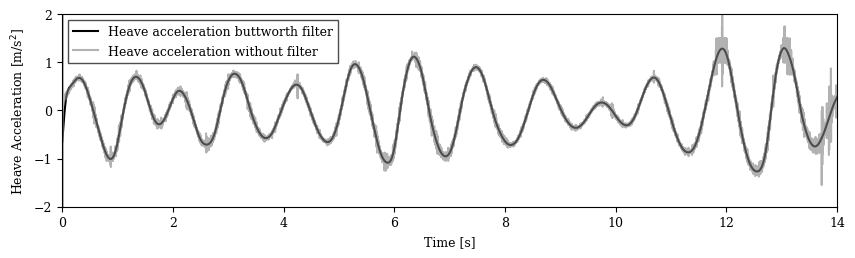}
\caption{Heave acceleration calculated from the unfiltered and filtered data. The dominant response is maintained, while high-frequency numerical noise is reduced without introducing an evident bias.}
\label{fig:filtering_accel}
\end{figure}

The raw surge, heave and pitch signals were preprocessed prior to system identification to reduce the amplification of high-frequency numerical noise during differentiation. The surge displacement \(x(t)\), heave displacement \(z(t)\), and pitch angle \(\theta(t)\) were filtered using a second-order, zero-phase Butterworth low-pass filter with a cutoff frequency of \(f_c=2.0~\text{Hz}\). Figure~\ref{fig:filtering_accel} demonstrates that the selected cutoff preserves the dominant acceleration response while eliminating the high-frequency fluctuations present in the state derivatives.

This preprocessing step is particularly important for SINDy, since errors in the estimated velocity and acceleration signals directly affect the identified model coefficients. In the present application, the dominant rigid-body and wave-induced motions occur at comparatively low frequencies, while the higher-frequency content is predominantly associated with numerical noise. This makes low-pass filtering appropriate for extracting the hydrodynamically relevant response prior to differentiation. 

After filtering, time derivatives were computed numerically using a second-order central-difference approximation,
\begin{equation}
\dot{q}_i \approx \frac{q_{i+1}-q_{i-1}}{2\Delta t}
\label{eq:dx}
\end{equation}
to obtain  \(\dot{x}\), \(\dot{z}\), \(\dot{\theta}\), and the wave-elevation derivatives \(\dot{\eta}_1\), \(\dot{\eta}_2\). Second, derivatives were obtained using the second-order central-difference formula,
\begin{equation}
\ddot{q}_i \approx \frac{q_{i+1}-2q_i+q_{i-1}}{\Delta t^2}
\label{eq:ddx}
\end{equation}
 to obtain the accelerations \(\ddot{q}_i\).

\subsection{System Identification}
\label{sec:sys-id}
System identification was done with constrained least squares regression with ridge regression. 

Given a matrix-library of terms \({\Theta}_i \in \mathbb{R}^{h_\mathrm{train}\times N_{i,terms}}\) and target derivative vector \({\ddot{s}_i} \in\mathbb{R}^{h_\mathrm{train}}\), where ${h_\mathrm{train}}$ is the number of time steps within the training horizon and ${N_{i,\mathrm{terms}}}$ is the number of terms, the coefficients contained in the matrix ${\Xi}_i \in \mathbb{R}^{{N_{i,terms}}}$ are estimated using a constrained least squares regression with ridge regulations such that

\begin{equation}
\arg\min_{{\Xi}_i}
\left\|
\begin{bmatrix}
{\Theta}_i\\
\sqrt{\alpha}\,I
\end{bmatrix}
{\Xi}_i
-
\begin{bmatrix}
{\ddot{s}}_i\\
{0}
\end{bmatrix}
\right\|_2^2
\quad
\text{subject to}
\quad
{lb}\le {\Xi}_i\le {ub},
\label{eq:bounded-ridge}
\end{equation}

Unlike standard ridge regression which has a closed form solution,
\begin{equation}
    \Xi = (\Theta^T\Theta + \alpha I)^{-1}\Theta^T \ddot{\mathbf{s}}
\end{equation}
bounded ridge regression has no such general closed form solution. Hence Eq. \ref{eq:bounded-ridge} is solved by solving a convex optimisation problem with an interior point-like method called 'Trust Region Reflective'. 

To ensure identified coefficients remain physical, bounds were imposed during each ridge-regression. In particular, restoring and damping terms were constrained to have non-positive coefficients, forcing terms are constrained for the direction of forcing, while all remaining terms are left unconstrained. The following bounds were imposed for surge Eq: \ref{eq:constraints_x}, heave Eq: \ref{eq:constraints_fz} and pitch Eq: \ref{eq:constraints_mtheta}.

\begin{equation}
\begin{aligned}
\xi_{\ddot{x},\dot{x}} &\le 0, &
\xi_{\ddot{x},|\dot{x}|\dot{x}} &\le 0,&
\xi_{\ddot{x},y_x} &\le 0, &
\xi_{\ddot{x},\eta_1} &\ge 0, &
\xi_{\ddot{x},\dot{\eta}_1} &\ge 0,&
\xi_{\ddot{x},\eta_2} &\le 0, &
\xi_{\ddot{x},\dot{\eta}_2} &\le 0
\end{aligned}
\label{eq:constraints_x}
\end{equation}

\begin{equation}
\begin{aligned}
\xi_{\ddot{z},z} &\le 0, &
\xi_{\ddot{z},z^3} &\le 0, &
\xi_{\ddot{z},\dot{z}} &\le 0, &
\xi_{\ddot{z},|\dot{z}|\dot{z}} &\le 0,&
\xi_{\ddot{z},\eta_1} &\ge 0, &
\xi_{\ddot{z},\eta_2} &\ge 0, &
\xi_{\ddot{z},\dot{\eta}_1} &\ge 0, &
\xi_{\ddot{z},\dot{\eta}_2} &\ge 0,&
\xi_{\ddot{z},y_{zz}} &\le 0
\end{aligned}
\label{eq:constraints_fz}
\end{equation}

\begin{equation}
\begin{aligned}
\xi_{\ddot{\theta},\theta} &\le 0, &
\xi_{\ddot{\theta},\theta^3} &\le 0,&
\xi_{\ddot{\theta},\dot{\theta}} &\le 0, &
\xi_{\ddot{\theta},|\dot{\theta}|\dot{\theta}} &\le 0,&
\xi_{\ddot{\theta},y_{\theta\theta}} &\le 0
\end{aligned}
\label{eq:constraints_mtheta}
\end{equation}

It was found that training based on approximately 8 characteristic wave periods yielded a robust result for future predictions in the equivalent sea state. Model training for both the linear and the nonlinear library takes on the order of $10^{-1}-10^0$ s to complete.

\subsubsection{Linear Library}
The linear system was identified where $\Theta_i$ only contains linear terms in state space and forcing in the library $\Theta_i$ in \ref{eq:surge-lin} \ref{eq:heave-lin} and \ref{eq:pitch-lin}. Heave and pitch are coupled whilst the dynamics in surge is independent. 
\begin{equation}
    {\Theta}_{\ddot{x}} =
        \begin{bmatrix}
        1,\,
        {\dot{x}},\,
        {\eta_1},\,
        {\dot{\eta}_1},\,
        {\eta_2},\,
        {\dot{\eta}_2},\,
        {y_x},\,
        {w_x}
        \end{bmatrix}
\label{eq:surge-lin}
\end{equation}
\begin{equation}
   {\Theta}_{\ddot{z}} =
        \begin{bmatrix}
        1,\,
        z,\,
        \dot{z},\,
        \eta_1,\,
        \dot{\eta_1},\,
        \eta_2,\,
        \dot{\eta_2},\,
        y_{zz},\,
        w_{zz},\,
        y_{z \theta},\,
        w_{z \theta},\,
        \theta,\, \dot{\theta}
        \end{bmatrix} 
        \label{eq:heave-lin}
\end{equation}

\begin{equation}
    {\Theta}_{\ddot{\theta}} =
        \begin{bmatrix}
        1,\,
        \theta,\, z,\,
        \dot{z},\,
        \dot{\theta},\,
        \eta_1,\,
        \dot{\eta}_1,\,
        \eta_2,\,
        \dot{\eta}_2,\,
        y_{\theta z},\,
        w_{\theta z},\,
        y_{\theta \theta},\,
        w_{\theta \theta},\,
        \end{bmatrix} 
        \label{eq:pitch-lin}
\end{equation}

\subsubsection{SINDy}
For a surrogate model identified with SINDy, a central assumption is that the governing dynamics can be represented using only a small subset of terms from a larger candidate library $\Theta$ of physically motivated basis terms.

Separate libraries are constructed for surge \ref{eq:surge-library}, heave \ref{eq:heave-library}, and pitch \ref{eq:pitch-library} accelerations. Rather than considering a full second order polynomial expansion of terms characteristic of a traditional SINDy approach, terms are only selected that represent interpretable coefficients relating to hydrostatics, viscous damping, wave forcing, and wave radiation.
The intentional selection of terms was found to be of key importance to reducing the effect of multi-colinearity and preventing solution instability outside of the training data.

\begin{equation}
    \bar{\Theta}_{\ddot{x}} =
        \begin{bmatrix}
        1,\,
        {\dot{x}},\,
        {\eta_1},\,
        {\dot{\eta}_1},\,
        {\eta_2},\,
        {\dot{\eta}_2},\,
        {y_x},\,
        {w_x},\,
        \end{bmatrix} / ||\Theta_{\ddot{x},j}||_2
\label{eq:surge-library}
\end{equation}
\begin{equation}
   \bar{\Theta}_{\ddot{z}} =
        \begin{bmatrix}
        1,\,
        z,\,
        \dot{z},\,
        |\dot{z}|\dot{z},\,
        \eta_1,\,
        \dot{\eta_1},\,
        z\eta_1,\,
        \dot{z}\eta_1,\,
        \eta_2,\,
        \dot{\eta_2},\,
        z\eta_2,\,
        \dot{z}\eta_2,\,
        y_{zz},\,
        w_{zz},\,
        y_{z \theta},\,
        w_{z \theta},\,
        \theta,\, \dot{\theta},\, \theta \eta_1,\,\dot{\theta} \eta_1,\, \theta \eta_2 ,\, \dot{\theta} \eta_2
        \end{bmatrix} /||\Theta_{\ddot{z},j}||_2
        \label{eq:heave-library}
\end{equation}

\begin{equation}
    \bar{\Theta}_{\ddot{\theta}} =
        \begin{bmatrix}
        1,\,
        \theta,\, z,\,
        \dot{z},\, z\theta,\, \dot{z}{\dot{\theta},\,}
        \theta^2,\,
        \theta^3,\,
        \dot{\theta},\,
        |\dot{z}|\dot{z},\,
        |\dot{\theta}|\dot{\theta},\,
        \eta_1,\,
        \dot{\eta}_1,\,
        \theta\eta_1,\,
        \dot{\theta}\eta_1,\,
        \eta_2,\,
        \dot{\eta}_2,\,
        \theta\eta_2,\,
        \dot{\theta}\eta_2,\,
        y_{\theta z},\,
        w_{\theta z},\,
        y_{\theta \theta},\,
        w_{\theta \theta},\,
        \end{bmatrix} / ||\Theta_{\ddot{\theta},j}||_2
        \label{eq:pitch-library}
\end{equation}

For SINDy, the matrix-library $\bar{\Theta}_i$ consists of normalised potential candidate terms which are automatically identified as contributing or 'not' contributing to the dynamics through the sequential least squares threshold STLSQ procedure. STLSQ is done to enforce model sparsity. Beginning with the full active set of candidate terms, coefficients smaller than a threshold are removed after each bounded ridge-regression solve as in Equation \ref{eq:bounded-ridge}. The magnitude of the thresholding value is ultimately determined by the hyperparameter optimisation described in Section \ref{sec:Hyperamater-optimisation}.

Such thresholding only makes sense if the fluctuations of the candidate library are of a comparable scale. If the fluctuations are of a comparable scale, then the magnitude of the identified normalized coefficients will be proportional to their importance in driving the dynamics. Therefore, before thresholding, each of the columns $j$ of the basis library $\Theta$ matrix are independently normalised by the $L_2$ norm as in Eq. \ref{eq:normalisation}:

\begin{equation}
    \bar{\Theta}_{:,j}=\frac{\Theta_{:,j}}{||\Theta_{:,j}||_2}
    \label{eq:normalisation}
\end{equation}

\subsection{Hyperparameter Optimisation}
\label{sec:Hyperamater-optimisation}
For the very first model training, the initial wave field is integrated forward in time with the PSPF solver up to the training horizon $h_{\mathrm{train}}$. The body's known x-coordinates $x_{\mathrm{true}}$ are used to determine the magnitude of the free-surface elevation at the virtual probe locations ($\eta_1(x_{\mathrm{true}}-0.5b)$, $\eta_2(x_{\mathrm{true}}+0.5b)$) that are incorporated into the library of candidate terms $\Theta_i$. 

The system identification hyperparameters were optimised using the Tree-structured Parzen Estimator (TPE) algorithm. TPE was selected as a sample-efficient, derivative-free Bayesian optimisation method suitable for the present high-dimensional non-convex hyperparameter optimisation problem \cite{Watanabe2023}.

For the linear only library the optimised hyperparameters comprise the decay and oscillation rates associated with the radiation-memory states $(\phi_x,\omega_x,\phi_{zz},\omega_{zz},\phi_{z\theta},\omega_{z\theta},\phi_{\theta z},\omega_{\theta z},\phi_{\theta\theta},\omega
_{\theta\theta})$, whilst the SINDy model also includes the sparsity thresholds in the optimisation loop. 

The objective function evaluates the accuracy of the integrated model response in surge, heave, and pitch while additionally penalising model complexity;

\begin{align}
    \mathcal{J}
=
\sum_{i=1}^{3}
\left[
\operatorname{nrmse}\!\left(s_i,s_{i,\mathrm{predict}}\right)
+
\operatorname{nrmse}\!\left(\dot{s}_i,\dot{s}_{i,\mathrm{predict}}\right)
+
\operatorname{nrmse}\!\left(\ddot{s}_i,\ddot{s}_{i,\mathrm{predict}}\right)
\right]
+
10^{-3}N_{i,\mathrm{terms}}, \\
\qquad
(s_1,s_2,s_3)=(x,z,\theta) \nonumber
\end{align}

Here, $N_{\mathrm{terms}}$ is the number of active terms identified with SINDy, which penalises higher complexity models under the assumptions that such models are likely to be less numerically stable and less robust. The normalised root-mean-square error is defined as
\begin{equation}
\operatorname{nrmse}(q,q_{\mathrm{predict}})
=
\frac{
\left\|q-q_{\mathrm{predict}}\right\|_2
}{
\left\|q\right\|_2+\epsilon
}
\end{equation}
where $\epsilon$ is a small positive constant used to avoid division by zero.

\begin{figure}
    \centering
    \includegraphics[width=0.9\linewidth]{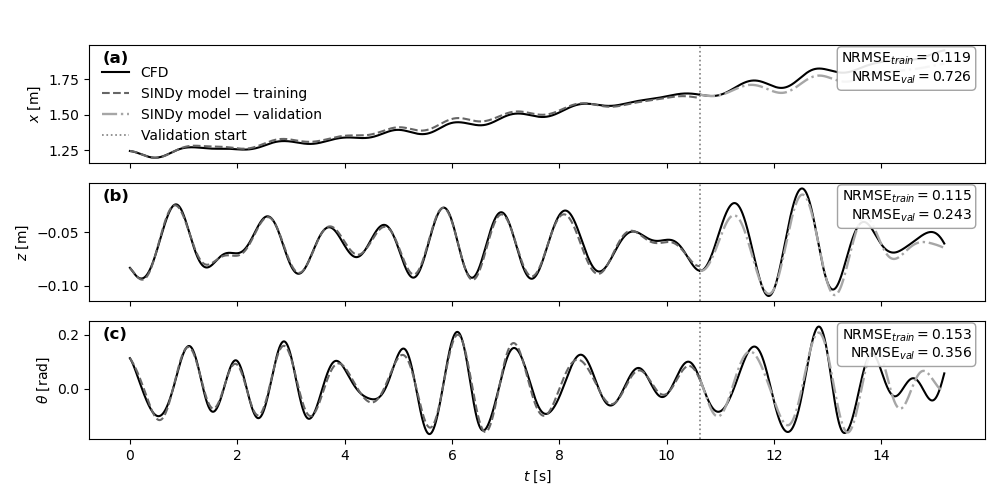}
    \caption{Training fits for the identified surrogate model: (a) surge, (b) heave, and (c) pitch.}
    \label{fig:regression-fits}
\end{figure}

Each hyperparameter trial requires both model identification and subsequent numerical integration of the identified system using RK4. Consequently, the objective function is expensive, with a complete optimisation comprising 50 trials requiring on the order of $10^1-10^2$ s to complete. Figure \ref{fig:regression-fits} shows the output after model training and hyperparmater optimisation. 

\begin{table}[]
\centering{}{
\begin{tabular}{c c c c c c c c}
\toprule
$\lambda_x$ &
$\lambda_{z}$ &
$\lambda_{\theta}$ &
$\sigma_x$ &
$\omega_x$ &
$\sigma_{zz}$ &
$\omega_{zz}$ \\
\midrule
0.3342 &
3.0313 &
3.1441 &
8.8961 &
17.4578 &
9.7965 &
5.6254 \\
\midrule
$\sigma_{z\theta}$ &
$\omega_{z \theta}$ &
$\sigma_{\theta z}$ &
$\omega_{\theta z}$ &
$\sigma_{\theta \theta}$ &
$\omega_{\theta \theta}$ &
\\
\midrule
3.9559 &
23.1626 &
4.6668 &
3.0511 &
13.6991 &
1.9446 &
\\
\bottomrule
\end{tabular}
}
\caption{Optimised sparsity thresholds and radiation-memory hyperparameters used in the identified surrogate model.}
\label{tab:optimized-hyperparameters}
\end{table}

Following optimisation, the hyperparameters are saved independently of the identified coefficients. This allows the optimised hyperparameter set to be reused during subsequent online model identification, avoiding the need to repeat the full expensive hyperparameter optimisation online during online model re-identification described in Section \ref{sec:online-training}.

\section{Online Implementation}

The identified dynamic models are coupled online with the PSPF solver through a receding-horizon prediction architecture. The purpose of this framework is to continuously provide a grey-box, accurate, deterministic medium-term prediction of the future vessel states capable of providing predictions in real-time applications.

\subsection{Prediction-Horizon Cycle Initalisation}
At the beginning of each prediction cycle, the most recent free-surface snapshot available at the current observation time, ($t_\mathrm{now}$), is extracted from the measured wave field and used to initialise the PSPF solver using the algorithm described in Eq \ref{eq:phis-from-eta}. The rigid-body dynamic model is simultaneously initialised using the corresponding measured state,
\begin{equation}
{s}_{\mathrm{true}}(t_\textrm{now})
=
\left[
x_{\mathrm{true}},\dot{x}_\mathrm{true},z_\mathrm{true},\dot{z}_\mathrm{true},\theta_\mathrm{true},\dot{\theta}_\mathrm{true}
\right]
\end{equation}
The radiation-memory states cannot be obtained directly from the instantaneous rigid-body state and are therefore reconstructed by integrating the memory-state equations over the available measured velocity history preceding ($t_\mathrm{now}$). The resulting body and memory states provide the initial condition for the subsequent coupled dynamic equations--PSPF prediction.

\subsection{Coupling Online System Identification and PSPF}
The potential flow solver is advanced over a prescribed prediction horizon ($h_{\mathrm{pred}}$). As the PSPF solution evolves, each newly calculated free-surface field is supplied to a bi-linear interpolator that matches the time and spatial coordinates of the output of the PSPF and the dynamic model. The predicted incident-wave input vector supplied to the hydrodynamic model at the future time ($t$) is ${w}_\mathrm{pred-PSPF}(t)$.

At each future time $t$, the predicted body state $s_{i,\mathrm{pred}}(t-1)$, radiation-memory states $m_{i,\mathrm{pred}}(t-1)$, and wave quantities $w_{\mathrm{pred-PSPF}}(t)$ are assembled into the vector $\Theta_{i,\mathrm{pred}}$ of size $N_{\mathrm{terms}} \times 1$. The subsequent future states are obtained directly from the SINDy model $\Xi_i$;
\begin{equation}
    {\ddot{s}}_{i,pred}(t) = {f}
\left({s}_\mathrm{pred}(t-1),m_\mathrm{pred}(t-1), {w}_\mathrm{pred-PSPF}(x(t),t-1); {\Xi}_i
\right)=\Theta_{i,\mathrm{pred}} \Xi_{i}, \qquad
    t \in [t_\mathrm{now}, t_\mathrm{now}+h_{\mathrm{pred}}]
\end{equation}
where the system is advanced sequentially forward in time using RK4.

\subsection{Model Output}
The receding horizon predictions operate sequentially. While a new SINDy--PSPF forecast is being generated, the most recently completed prediction remains active and continues to provide the currently best future trajectory prediction. When a newly computed prediction becomes available, it immediately supersedes the previous prediction,
\begin{equation}
\mathcal{P}^{(k)}
\rightarrow
\mathcal{P}^{(k+1)}
\qquad
\text{when }
\mathcal{P}^{(k+1)}
\text{ becomes available}.
\end{equation}
Here, $\mathcal{P}$ is the future motion prediction and $(k)$ denotes the prediction-cycle index. The latest completed forecast is presented as the current best estimate of the future state, while the next prediction cycle is initiated from measured wave field and body state at the time the previous forecast becomes available $t_{\mathrm{now}}^{k+1}$. This receding-horizon process is repeated continuously throughout the online prediction period until the dataset no longer contains state or wave field data at the current time. 

\subsection{Online Model Updating}
\label{sec:online-training}
In addition to repeatedly updating the incident-wave forecast, the coefficient vectors $\Xi_i$ are periodically re-identified using a rolling window of known past data states and importantly, 

The measured true past body position, velocity, and acceleration histories over the training interval horizon $h_\mathrm{train}$ are combined with the saved PSPF predictions of the incident free-surface elevations at the true body's past $x-$locations whereby the PSPF was initialised at the start of the online training window. Futhermore, during training, the previously optimised hyperparameters are retained while the identified coefficients are allowed to adapt to changing wave conditions The training libraries may be written as
\begin{equation}
{\Theta}_{i}^{(v+1)}
=
{\Theta}_{i}
\left(
{s}_{\mathrm{true}}(t),m_{\mathrm{true}}(t),
{w}_{\mathrm{pred-PSPF}}
\bigl(
x_{\mathrm{true}}(t),t
\bigr)
\right),
\qquad
t\in \left[
t_{\mathrm{now}}-h_{\mathrm{train}},
t_{\mathrm{now}}
\right],
\end{equation}
with the corresponding measured acceleration targets
\begin{equation}
\ddot{s}_{i,true}^{(v+1)}(t),
\qquad
t\in\ \left[
t_\mathrm{now}-h_{\mathrm{train}},
t_\mathrm{now}
\right]
\end{equation}

The coefficients are re-identified online using bounded ridge regression. When a newly computed model becomes available, it superseeds the old model,
\begin{equation}
    \Xi_i^{(v)} \rightarrow \Xi_i^{(v+1)} \qquad
\text{when }
{\Xi}^{(v+1)}
\text{ becomes available}.
\end{equation}
where $\Xi_i$ is the surrogate model and $(v)$ is the active model version. Once complete, the coefficient vectors in the active model are replaced by the newly identified values. Subsequent forecasts are then generated using the updated coefficient set. 

\section{Results}
This section presents the results of the surrogate modelling framework in terms of computational speed, regression accuracy, forecast horizon, and robustness under online prediction.

At each prediction step, only the state and incident wave information available up to the current time is supplied to the model, while all subsequent CFD data are withheld from the prediction. The resulting forecasts therefore represent entirely out-of-sample predictions, as no future body states or wave elevations are used in the predictions.

\begin{figure}
    \centering
    \includegraphics[width=0.9\linewidth]{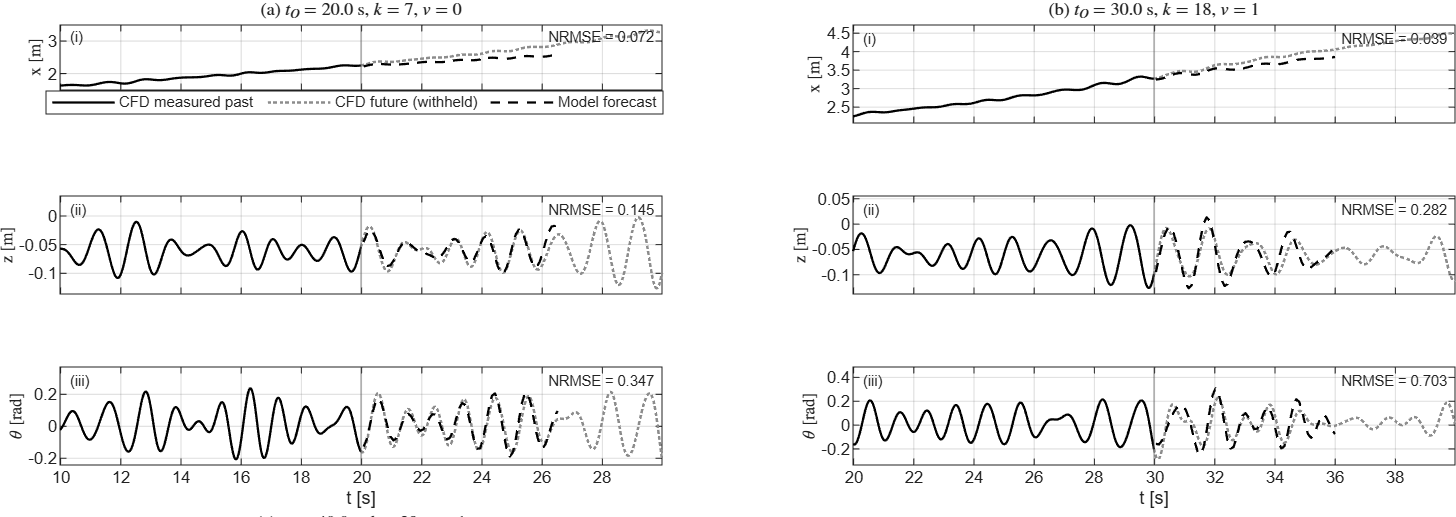}
    \caption{Online future motion prediction of linear only model at (a) 20s and (b) 30s in real time.}
    \label{fig:Online-future-motion-prediction-Lin}
\end{figure}

\begin{figure}
    \centering
    \includegraphics[width=0.9\linewidth]{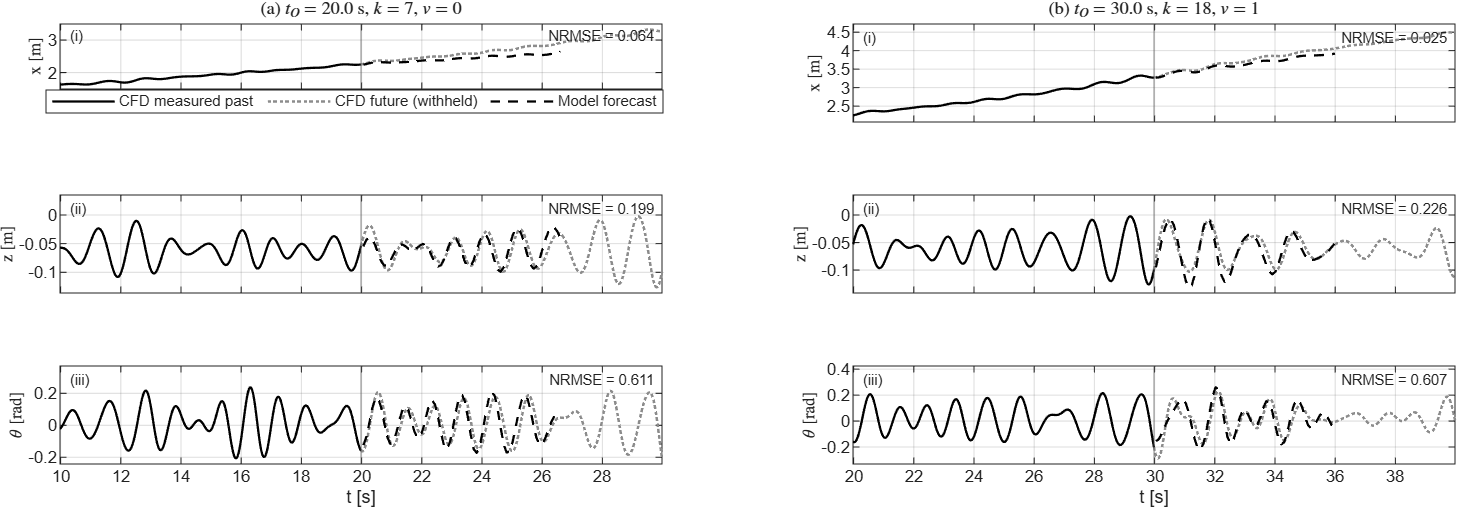}
    \caption{Online future motion prediction of SINDy model at (a) 10 s and (b) 20 s in real time.}
    \label{fig:Online-future-motion-prediction-SINDy}
\end{figure}

Figure~\ref{fig:Online-future-motion-prediction-Lin}(a)--(b) and Figure~\ref{fig:Online-future-motion-prediction-SINDy}(a)--(b) show the corresponding surge, heave, and pitch responses for the linear only model and the SINDy models respectively. The vertical line denotes the current prediction time, separating the state history available to the model from the future prediction interval. The most recent model forecast is shown by the black dotted line, while the grey dotted line represents the withheld CFD, against which, the forecast is assessed.

The index $k$ shown in each panel denotes the most recent prediction step, while the $v$ refers to the must up to date model coefficients, ${\Xi}$, used for that forecast. For each prediction, the NRMSE is evaluated from the current prediction time to the end of the forecast horizon. 

The average time it took to compute a 7.741 s future horizon prediction is 0.767 s meaning that it is only predicting 6.974 s into the future by the time it comes online. The average time it took to compute a new model is 1.529 s.

\begin{figure}
    \centering
    \includegraphics[width=1.0\linewidth]{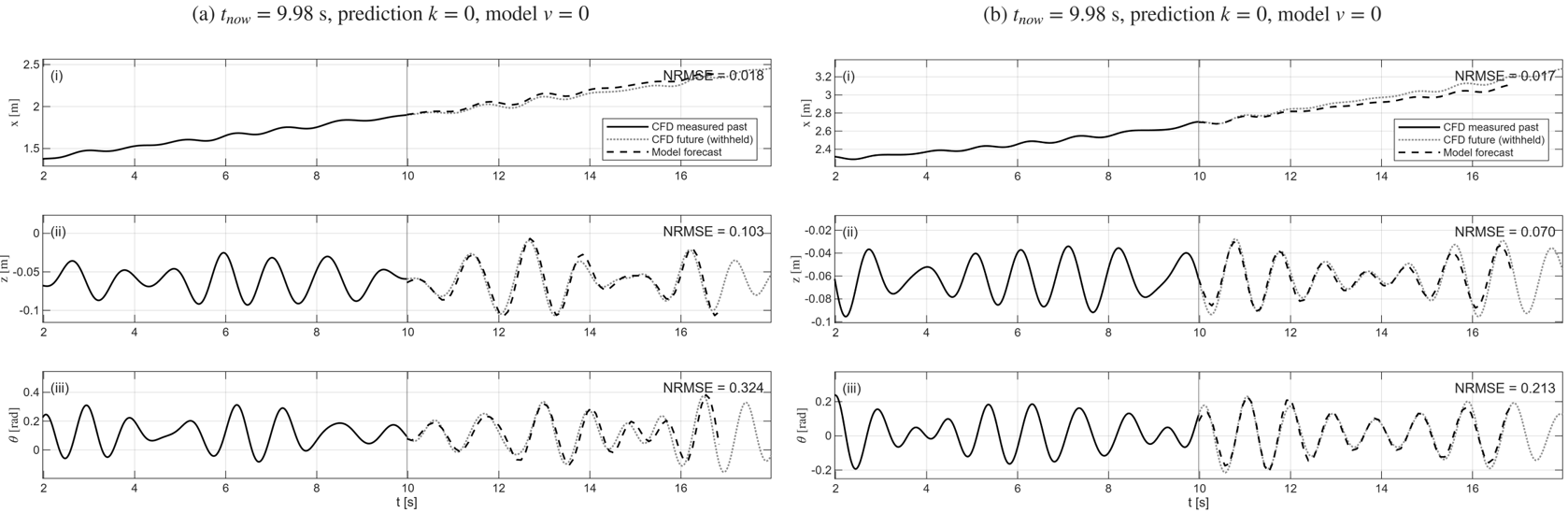}
    \caption{Online future motion prediction for (a) box of 0.75 width of reference box and (b) reference box with 50\% greater peak wave frequency.}
    \label{fig:diff-shape-spectra}
\end{figure}

Without any changes to the code, the framework is shown in Figure \ref{fig:diff-shape-spectra}(a) to perform equivalently well for a box geometry where the width is changed to 0.225 m and the inertial properties were changed accordingly. So also for the scenario where the peak period of incoming waves decreases as shown in Figures \ref{fig:diff-shape-spectra}(b). 

\section{Discussion}
The results confirm the effectiveness of separating out the the problem into an incident-wave prediction and a data-driven body dynamics model. Two key sources of error exist in this approach. The first source, is the difference between the true incident wave elevation at the fore and aft locations of the body against the PF prediction in Fig. \ref{fig:CFD-PF-comparison}. The second is the SINDy model error between the identified dynamic model and the true dynamic response in Fig \ref{fig:regression-fits}. An NRMSE of $0.058$ between the PSPF and CFD incident-wave elevations at the fore and aft body locations indicates that the wave-prediction component introduces a relatively small source of uncertainty into the forecast. The larger discrepancies in the final body-motion prediction is introduced from the identified dynamics model. 

The first updated model $\Xi_i^{v=1}$, utilising data from $10 \rightarrow18$ s of real-time past data states and PSPF predictions. For the SINDy model, only 31 of a potential 53 different basis terms are active after thresholding. It is evident that multi-colinearity, remains a key issue because multiple correlated candidate terms can represent similar dynamics, it is important to note that an identified coefficient is not automatically an independently identifiable physical hydrodynamic coefficient. This can be seen specifically for large opposite signed identified coefficients between $\eta_1$ and $\eta_2$. As these terms measure the wave elevation at nearby location on the box.  

The prediction behaviour differs between the three degrees of freedom. Heave and pitch are well behaved as they are bounded by restoring effects. Surge on the other hand has no physically equivalent restoring term. This makes the surge prediction particularly sensitive to small errors in the mean identified acceleration and velocity. Even a small offset in $\ddot{x}$ once integrated into displacement, produces a positional drift that progressively diverges from the reference CFD solution over the forecast horizon. The resulting error is also coupled back into the prediction because the incident-wave probes move with the predicted horizontal position of the body. An error in $x$ therefore not only affects the surge response itself, but also changes the locations at which $\eta_1$, $\eta_2$, $\dot{\eta}_1$ and $\dot{\eta}_2$ are sampled. In reality, for a self-propelled vessel, the surge motion is typically determined by speed-control systems, meaning that the future surge trajectory is constrained and the passive wave-induced surge dynamics of the case in this paper is less representative of a controlled ship application.

The stability of the numerical integration and system identification was found to be improved by constraining the candidate library to linear only terms. Given that both linear and nonlinear formulations achieved comparable predictive accuracy, the additional nonlinear terms were found to not provide an improvement under the conditions investigated. The linear formulation is therefore preferred for its greater numerical stability, reduced complexity and more consistent identification. We hypothesis that the ability for the linear only library to maintain predictive accuracy in the larger sea states and around resonance conditions where significant vortex shedding and nonlinear diffraction is observed can be attributed to the online re-identification. This may enable the linear coefficients to implicitly account for the influence of nonlinear hydrodynamic effects in locally similar wave regime where the recent sea-state history used for training does not significantly change from the waves observed over the future prediction horizon. 

Furthermore, the online implementation limits the accumulation of error in surge. Each newly completed forecast is reinitialised from the most recent measured body state and wave field, preventing errors from one prediction cycle from propagating into subsequent cycles. The useful prediction horizon is therefore governed by the interval over which the model remains sufficiently accurate.

The benefit of online re-identification is expected to become more significant when the operating condition changes. A fixed coefficient set assumes that the effective mapping between incident-wave conditions and body response remains adequately represented by the original identification data. This assumption becomes progressively weaker as wave frequency, steepness, or other operating conditions change. Updating the sparse coefficients provides a mechanism by which the model can adapt while retaining the same physically motivated candidate structure. However, if the underlying dynamics change such that physically relevant terms are absent from the original candidate library, changing the coefficients alone may not be able to recover the missing physics.

An important feature of the present formulation is that the computational cost is linked primarily to the number of wave periods that must be propagated rather than directly to dimensional time. The model-scale case considered here has a dominant period of approximately $1.2~\mathrm{s}$, whereas representative full-scale sea states have substantially longer periods. 

\section{Conclusion}
This study demonstrates a proposed surrogate modelling framework capable of deterministically predicting future motion in irregular waves in real-time. The CFD solution is used as the reference against which the forecast is assessed, but future CFD wave and body-motion data are withheld when each prediction is generated. The online results therefore demonstrate the out-of sample forecasting ability of the data driven model.

The present study considers a two-dimensional body in unidirectional irregular waves. Extension to a three-dimensional ship may incorporate directional wave propagation, additional rigid-body degrees of freedom and larger library $\Theta$ of candidate basis terms. Short-crested seas introduce a further challenge because the incident wave field can no longer be represented by a single horizontal coordinate. The PSPF forecast would require a three-dimensional domain, increasing both the number of spectral degrees of freedom and change the wave-field initialisation procedure.

A practical implementation also requires the initial free-surface field to be obtained from onboard measurements rather than CFD. The present PSPF initialisation is intentionally based only on surface elevation because this is the quantity that can be reconstructed using radar. Recovering a sufficiently accurate phase-resolved incident-wave state from radar therefore remains an important step between the present numerical demonstration and deployment at sea.

Overall, the results support the capacity of the proposed approach for accurate short term online deterministic motion forecasting. The current limitations relate primarily to accumulated surge error, the assumption of unidirectional seas, and the computational expense of evaluating the PSPF solver in faster-than-real time. Future work will focus on the extension of the identified structure to the three-dimensional case, improving efficiency of code implementation for computational speed improvements, model implementation for multiple body interactions, and implementation for short-crested/directional ocean waves.

\section*{Acknowledgments}
\begin{small}
The authors gratefully acknowledge the support of the CSIRO Data61 Next Generation Graduate Program (NGGP) and Navantia Australia for sponsoring this research. Their financial support, resources, and ongoing engagement have been instrumental in enabling the successful completion of this work. The authors also appreciate the valuable insights and collaboration provided through this partnership, which have significantly contributed to the development and outcomes of this study.
\end{small} 

\section*{Disclosure}
During the preparation of this report, the authors used ChatGPT (OpenAI), Claude (Anthropic), and Gemini (Google) to assist with code debugging. After using this tool/service, I reviewed and edited the content as needed and I take full responsibility for its content.


\bibliographystyle{cas-model2-names}
\bibliography{references2}

\end{document}